\documentstyle[aps,prd,epsfig,floats]{revtex} 

\draft

\begin{document}
\twocolumn[\hsize\textwidth\columnwidth\hsize\csname
@twocolumnfalse\endcsname
\title{%
\hbox to\hsize{\normalsize\rm 
\hfil Preprint MPI-PhT/2000-38}
\vskip 36pt     Adiabatic Inversion in the SQUID, Macroscopic
Coherence and Decoherence}

\author{P. Silvestrini}
\address{Instituto di Cibernetica del CNR, via Toiano 6, I-
80072, Arco Felice, Italy and
MQC group, INFN/Napoli, Italy}

\author{L.~Stodolsky}
\address{Max-Planck-Institut f\"ur Physik 
(Werner-Heisenberg-Institut),
F\"ohringer Ring 6, 80805 M\"unchen, Germany}

\maketitle

\centerline{              }
\centerline{ Presented by LS at the Conference on Macroscopic
Coherence and Quantum Computing}
\centerline {Naples, June 2000}

\begin{abstract}

  A procedure
for demonstrating quantum coherence and measuring decoherence times
between different
fluxoid states of a SQUID by using  ``adiabatic
inversion'' is discussed. One  fluxoid state is smoothly
transferred into
the other,  like a spin  reversing direction by following  a slowly
moving
 magnetic field. This is accomplished by sweeping an external
applied flux, and
  depends on a well-defined quantum
phase between the two macroscopic states.  Varying the speed of the
sweep relative to the decoherence time permits one to move from the
quantum regime, where such a well-defined phase exists, to the
classical regime where it is lost and
the inversion is inhibited. Thus  observing 
whether inversion
has taken place or not as a function of sweep speed
 offers the possibility of measuring
the  decoherence time.  Estimates with some  typical SQUID
parameters  are presented and it appears that such a procedure
should be experimentally possible. The main
 requirement for the feasibility of the scheme appears to be that
the low
temperature relaxation
time among the quantum levels of the SQUID be long compared to
other  time scales of the problem, including the readout time.
Applications to
 the ``quantum computer'', with the  level system
 of the SQUID playing the role  of the qbit, are briefly examined.
\end{abstract}
\vskip2.0pc]

\section{Introduction}
 One of the first and certainly probably the most discussed systems
in which one tries to demonstrate coherence between apparently
macroscopically different states is the
SQUID~\cite{leggett}.

 In the last decade a number of beautiful experiments at
low temperature~\cite{lukens} have seen effects connected with the
quantized  energy levels~\cite{clarke}
expected in the SQUID, showing that it does in fact in many ways
resemble a ``macroscopic atom''.
A fast sweeping
method~\cite{obs} has also
seen the effects of these quantized levels even at relatively high
temperature. 

 If we could further  show in some way that the quantum phase
between two fluxoid states of the SQUID, where a great number of
electrons go around a ring in one direction or the other, is
physically meaningful and observable it would  certainly provide
even more impressive evidence for the applicability of quantum
mechanics on large scales and  help put to sleep any ideas about
the existence of some mysterious scale where quantum mechanics
stops working.

 At this meeting,   results using microwave spectroscopy have
indeed  been  reported   showing the repulsion of energy levels 
expected from the  quantum
mechanical mixing  of different fluxoid states \cite{luk1}. These
results indicate by implication that two opposite fluxoid states
can exist in meaningful linear combinations and so that their 
relative phase is significant.

 Here  we would like to propose another method for demonstrating
the   coherence between opposite fluxoid states and the
meaningfulness of the phase between them. Furthermore the method to
be described  allows a measurement of the ``decoherence time'', 
the  time in which the definite phase or the coherence between the
two states is lost. This is interesting of itself since it never
has been done and is also relevant to the ``quantum computer''
where decoherence is the main  obstacle to be overcome.

 Our basic idea  is to use the process of  
adiabatic inversion or level crossing where a slowly varying field
is used to reverse the states of a quantum
system.
In its most straightforward realization, our proposal consists of
starting with the SQUID in its lowest state, making a fast but
adiabatic sweep, and reading out to see if the final state is the
same or the opposite fluxoid state. If the state has switched to
the other fluxoid, the system has behaved quantum mechanically,
with phase coherence between the two states. If it stays in the
original fluxoid state, the phase coherence between the states was
lost and the system  behaved classically (see Figs. 3 and 4).
 
 In order for these effects to be unambiguously observable,
relaxation towards thermal equilibrium must be small on the time
scales involved in the experiment. As is discussed in section IV,
this seems to be obtainable  at low temperature  with suitably
chosen sweep times.

A very interesting aspect of the present proposal, as we shall
discuss
below, is the possiblity of passing between these two regimes, 
quantum and  classical, by simply varying 
the sweep speed. This allows us to obtain, for adiabatic
conditions,  the decoherence
time as the longest sweep time for which
the inversion is successful.

\section{Adiabatic Inversion in the rf Squid}
 
Most workers in this field are probably accustomed to
 discussing  level crossing problems such as ours in a Landau-Zener
picture where one exhibits the spacing of  the two crossing  energy
levels as some external parameter or field is varied. This is of
course a perfectly good and useful picture for many
applications. However I would like to urge consideration of another
perhaps more intuitive visualization, one which is particularly
well suited to time dependent problems as we have here, and which
also has the
advantage that it gives a complete representation of the  state of
the system at a given time and not just the level splitting.

This picture utilizes the fact that any
  two-state system  may viewed as constituting the two
components of
a ``spin''. Looking at this ``spin'' and its
 motions  then provides an easy visualization, one which has been
used in many contexts~\cite{us}. That is, if we have two states
$\vert L>$ and $\vert R>$, which  for us will be the two lowest
opposing fluxoid states, then the general state 
\begin{equation}\label{lin}
 \alpha \vert L>+\beta \vert R\rangle
\end{equation}
has the interpretation of a spin pointing in some direction. Even
the relative phase is represented: the numbers $\alpha$ and $\beta$
are complex numbers and the spin points in different directions
when we change their relative phase. Thus the spin visualizastion
gives a complete picture of the state (up to an irrelevant overall
phase).

To identify $\vert L>$ and $\vert R>$ here, consider (Fig. 1)    
the   familiar  double  well potential ~\cite{barone} for the rf
SQUID biased with an external flux $\Phi_x$
\begin{equation}
\label{u}
U=U_0[1/2 (\Phi-\Phi_x)^2 + \beta_L cos\Phi]\;.
\end{equation}
The horizontal axis, the ``position'' coordinate of the present
problem, is $\Phi$ the flux in  the SQUID ring.
 Depending on whether the system is in the left or right well, the
flux through the ring has a different sign and  the current goes
around
 in opposite directions.
   The potential can be varied by altering the
external $\Phi_x$,  becoming 
symmetric for $\Phi_x=0$. We identify the lowest level in each well
with $\vert L>$ and $\vert R>$.  Quantum mechanical linear
combinations of them result from tunneling through the barrier.  In
the adiabatic inversion procedure to be discussed, the external 
$\Phi_x$ is swept from a maximum to an minimum value, passing
through zero, such that the
initially
asymmetric left and right wells exchange roles, the
originally higher
well becoming the lower one and vice versa. The asymmetry of the
configurations,
 however, is kept small so that we effectively have only
a two-state 
system, composed of the lowest state in each well. 

\begin{figure}[h]
\epsfig{file=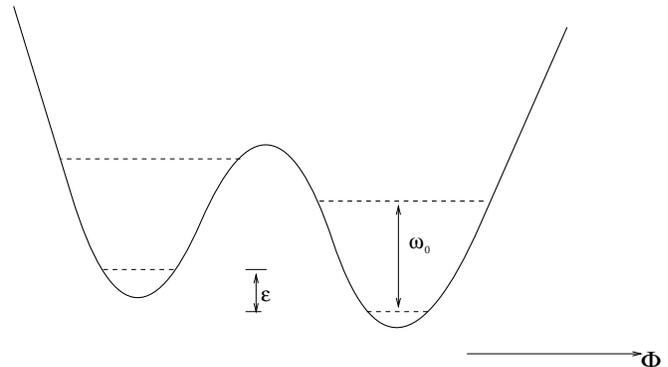, width=\hsize}
\caption{ Double potential well with harmonic level spacing 
$\omega_0$ and initial spacing $\epsilon$ between lowest levels.}
\end{figure}

 Various influences   such as the tunneling energy or the  external
flux may affect the orientation of the spin.
 Pursuing the  analogy, these influences may be thought of as
creating a kind of  pseudo-magnetic field $\bf V$  which causes 
the spin to precess. Representing the spin by a  ``polarization''
$\bf P$ we get the picture of 
Fig. 2. The equation governing the motion of $\bf P$ is 

\begin{equation}\label{pdot}
{\bf \dot P} = {\bf V} \times {\bf P} -D {\bf P}_{tr}\; ,
\end{equation}
where $\bf V$ can be time dependent. The quantity $D$ is the
decoherence parameter, which 
 we  neglect for the moment and will  deal with below.
 Note that in the absence of $D$,  according to Eq [\ref{pdot}],
$\bf P$ cannot change its length,
so the density matrix, which  $\bf P$ parameterizes, retains its
degree of purity.  In using Eq [\ref{pdot}]
 one assumes the temperature low enough so that relaxation
 processes like barrier hoping or jumps between prinicpal levels
may be neglected, thus there is
 damping of  
  only the transverse components,  ${\bf P}_{tr}$ (see below).

 Now it is a familiar fact  under adiabatic conditions, where $\bf
V$
varies slowly, that the  ``spin'' $\bf P$
will tend to ``follow'' a moving magnetic field ${\bf V}(t)$.   
This is a completely familiar procedure  when rotating the spin of
say an atom or
 a neutron by a magnetic field. If we wish, we can completely
invert the state by having $\bf V$ swing from ``up'' to ``down''.

\begin{figure}[h] 
\epsfig{file=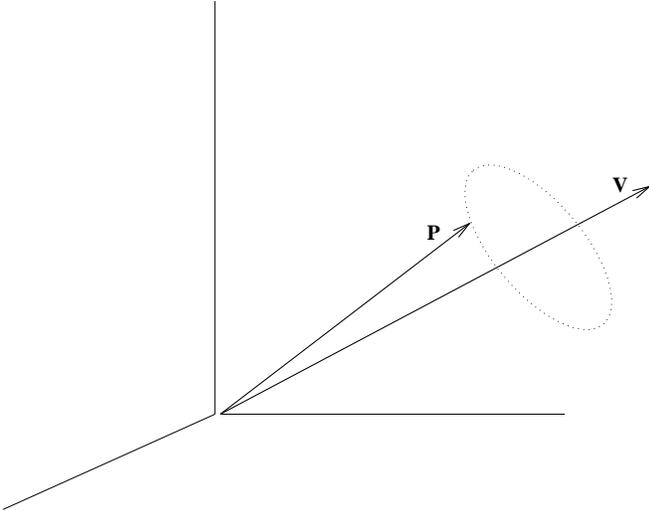, width=\hsize}
\caption{ Precession of the ``spin'' vector $\bf P$
 around the pseudo-magnetic field vector ${\bf V}(t)$. In adiabatic
inversion $\bf V$  swings from ``up'' to
``down'' and carries $\bf P$ with it.}
\end{figure}
In the present problem, we can create a moving
 $\bf V$ by sweeping $\Phi_x$. This is   because the vertical
component of $\bf V$, $V _{vert}$
corresponds to the  difference in the two lowest energy
levels. Hence we can induce a level crossing and reverse the
potential wells (see Figs. 3 and 4) by reversing $\Phi_x$ and so
the
direction of $V_{vert}$. For small asymmetry  of the wells the
splitting is linear in $\Phi_x$. Thus if $\epsilon$ is the initial
level splitting, obtained when $\Phi_x=\Phi_x^{max}$, we can write
$V_{vert}(t)=\epsilon (\Phi_x(t)/\Phi_x^{max})$.  As $\Phi_x$
sweeps from its positive maximum value to its negative minimum
value,
$V_{vert}$ reverses direction.

 In doing so, $V_{vert}$ passes through zero where $\bf V$ is
horizontal with only a transverse component $\bf V_{tr}$.  
 ${\bf V}_{tr}$ corresponds to the tunneling energy
between
the two quasi-degenerate states, $V_{tr}=\omega_{tunnel}$. As
$V_{vert}$  passes through zero at 
$\Phi_x=0$, the $\vert L>$ and $\vert R>$ states are strongly mixed
and the splitting of the resulting energy eigenstates is determined
by  $V_{tr}$ alone, as is also familiar in the Landau-Zener
picture.  The magnitude of $V(t)$ at a given time,  
$|{\bf V}|=\sqrt{V_{vert}^2+V_{tr}^2}$  gives the instantaneous
splitting of the two levels. This varies from approximately
$\epsilon$ in the vertical position of $\bf V$ to
$\omega_{tunnel}$ in the horizontal position.  

Having identified the components of $\bf V$, our next task is to
ascertain  the meaning of
``adiabatic'' or ``slow'' for the motion of $\bf V$. Adiabatic 
conditions obtain when the time variation in question  does not 
contain significant frequencies or fourier components corresponding
to the energy splitting between levels.  Expressed in terms of
time, the rate of variation of  $\bf V$
should be slow on the time scale corresponding to the tunneling
time between the two states. Thus we have the requirement on $\bf
V$ that its relative rate of variation $\dot V/V$ always be small
compared to $ V$ itself. Since the varying component of $\bf V$ is
$V_{vertical}$ (neglecting the indirect effect of $\Phi_x$ on
$\omega_{tunnel}$) we require  $\dot V_{vertical}/V<<V $. Thus
taking the near degenerate configuration where $V\approx
\omega_{tunnel} $ we find  
\begin{equation}\label{adb}
 \epsilon {\dot \Phi_x(t)\over \Phi_x^{max}}\approx \epsilon
\omega_{sweep} << \omega_{tunnel}^2
\end{equation}
 as the
condition
 for adiabaticity.

 If the adiabatic condition is violated then $\bf P$ cannot follow
$\bf V $ (See Fig. 6 below) .
We stress that for adiabatic inversion there must be a
 well-defined quantum phase between the two states, and that when
this
phase is lost the inversion is suppressed.

\begin{figure}[h]
\epsfig{file=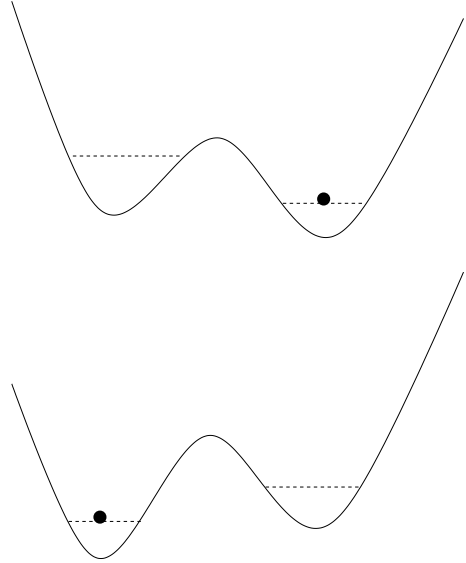, width=.7\hsize}
\hskip.3cm
\caption{ A succesful inversion, starting from the upper figure and
ending with the lower figure. The black dot indicates which state
is occupied. The system starts in the lowest energy level and stays
there, reversing fluxoid states. It behaves as a quantum system
with definite phase relations between the two states. }
\end{figure}

\begin{figure}[h]
\epsfig{file=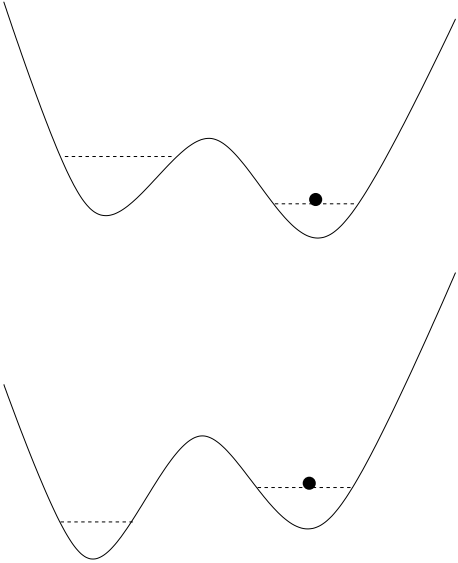, width=.7\hsize}
\caption{ An inhibited inversion, starting from the upper figure
and ending with the lower figure. Due to the lack of phase
coherence the system
 behaves classically and  stays in the same state, the flux  is not
reversed.} \end{figure}

In
 Figs. 3 and  4  we give a schematic representation of the whole
procedure. In Fig. 3 a succesful inversion takes place.
Starting (upper sketch) with the system in  the lowest state since
we are at low temperature, a sweep is performed. After the level
crossing has been performed (lower sketch ), the system has
reversed flux, remaining in the lowest energy state. In the
visualization of Fig. 2,  $ P_{vert}$ has reversed direction. This
is the behavior to be expected of a quantum mechanical system with
well-defined phase relations.

In Fig. 4 we have the same starting situation but the inversion  is
unsuccessful. The system ends up in the same fluxoid state, $
P_{vert}$ does not reverse. This is the behavior to be expected
classically, when decoherence is significant and there is no well-
defined quantum phase between the two states.

  More detailed  information, as well as intermediate cases will be
discussed in terms of the behavior of $\bf P$, as shown in Fig. 5
and Fig. 6 below. For this a  discussion of the role of damping or
decoherence is necessary. This is the topic of the next section.

\section{Damping}
 We now turn to a quantitative discusssion of 
dissipative or damping effects, those effects 
tending to destroy the quantum coherence of the system; that is  we
consider the role of the  $D$ term in Eq [\ref{pdot}]. While the
vertical component  of $\bf P$ characterizes the relative amounts
of the two states which are present in a probablistic sense, the
quantity
${\bf P}_{tr}$
 measures  the degree of phase coherence between the
two states. $D$ gives the rate of loss of this phase coherence.

One could also have included a damping parameter for the vertical
component of ${\bf P}$.  Instead of the loss of phase coherence,
this would  represent direct transitions from one well to the
other.  Such relaxation processes, like
 jumping the potential barrier due to thermal effects, or
``radiative transitions''  from one well to another with emission
or absorption of  some energy  can be minimized by low temperatures
and fast sweep times, as discussed in the next section. 

In the present problem 
 we are in  particular interested in the effect of $D$ on the 
inversion process.  With an increasing loss of phase
coherence, we expect the situation to become more  and more
``classical'' and finally when the $D$ is large, for the  
inversion to be inhibited. Indeed, in solving Eq[\ref{pdot}] in the
limit of large $D$ one finds the inversion is strongly blocked and 
that
 one arrives in the ``Turing-Watched Pot-Zeno'' regime
where ${\bf P}_{vert} $ is essentially ``frozen"~\cite{us}. (See
the lower right panels of Figs. 5 and 6).

 These general  expectations are confirmed by a numerical
study~\cite{flaig} of
  Eq [\ref{pdot}] with a  moving $\bf V(t)$.
 Figs. 5 and 6  show  results of the numerical study. The
horizontal
axis represents the time, running  from the beginning to the end of
the sweep. 
The two curves in each  picture represent   ${\bf P}_{vert} $
(labeled $P_3$) and ${\bf P}_{trans}$. ${\bf P}_{trans}$ is
represented in absolute value, while  ${\bf P}_{vert} $ can change
sign.   In Fig. 5 an adiabatic sweep is performed. In the first
panel,
where $D=0$, one sees the succesful inversion of ${\bf P}_{vert} $
as it moves from +1 to -1. ${\bf P}_{trans}$ passes through 1 as 
${\bf P}$ passes through the horizontal position. In the next
panel, where $D$ is increased somewhat, to $0.01$, the picture is
essentially
unchanged. In the third panel where $D= 0.2$, decoherence takes
effect since the decoherence time, given by $1/D$, is on the order
of the sweep time. ${\bf P}$ shrinks to zero during the sweep,
showing the loss of phase coherence.  Finally, with very large $D$,
${\bf P}_{vert} $ evolves hardly at all during the sweep and ${\bf
P}_{trans}$ stays at zero. This final case is the ``Turing-Watched
Pot-Zeno'' behavior.

 We see  three regimes for the result of the sweep: A) the system
stays in the original state (classical case), B) goes to the
opposite state (quantum case), or C) ends sometimes in one state
and sometimes in the other
(intermediate case). These correspond to $P_{vert}$ starting as +1
and ending as  +1, -1, or 0, respectively. Or in terms of Fig. 5,
A)
corresponds to the last panel (lower right), B) to the first two
panels, and C) to the middle panel (lower left).

 We stress that all this arises simply from solving  Eq
[\ref{pdot}] and that no further notions or assumptions are
necessary. 

\begin{figure}[h] 
\epsfig{file=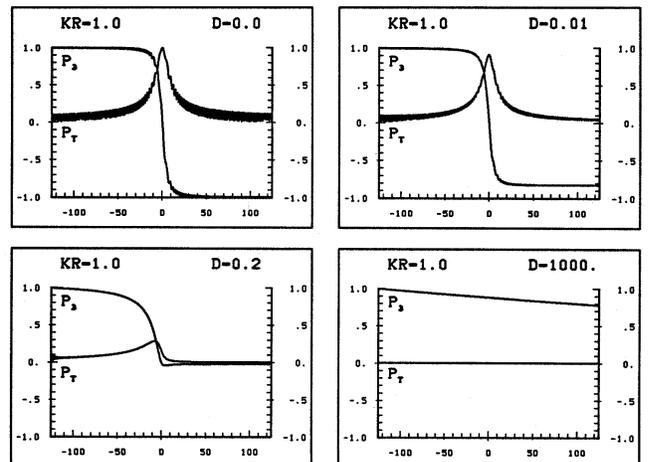, width=\hsize}
\caption{ The vertical and horizontal components of the vector $P$ 
as a function of time, for a sweep of the psuedofield $V_{vert}$
(not shown).  The four panels show increasing
values of  the damping
paramater D.    Adiabatic inversion is seen to occur for $D=0$ and
$D=.01$,   while further increasing D
blocks the inversion. For very large $D$  the ``Turing-Watched Pot-
Zeno''behavior sets in, where $P_{vert}$ evolves extremely slowly.
The curve starting at 1 is $P_{vert}$ and the curve starting at
zero is $P_{tr}$.  The time and $1/D$ are  in units of $1/V_{tr} =
1/\omega_{tunnel}$.  An  adiabaticity parameter
($KR$) is kept at a moderate value corresponding, in these units,
to a sweep time of about 20.}
\end{figure}

In Fig. 6 we show the effects of non-adiabaticity, using a sweep a
factor of ten faster than in Fig. 5. Even with $D=0$  the inversion
is incomplete, with $\bf P$ keeping unit length, but ending up
rotating roughly in the horizontal plane.  As $D$ is turned on,
$\bf P$ shrinks and  finally for very large $D$  the ``Turing-
Watched Pot-Zeno''  behavior sets in.

  Given adiabatic conditions, the important
relation is that between the sweep speed
and the decoherence time $1/D$. As one sees from Fig. 5, there is
no effective decoherence  until $1/D$ is on the order of the sweep
time. This relation  determines if the system has time to
``decohere'' during the sweep. Roughly speaking, one enters the
classical regime when $P_{tr}$, characterizing the phase coherence,
has time to shrink during the inversion.  
 This is very interesting for us,
 since  it means, if the experimental situation is
favorable, that we can  pass from the quantum regime to the
classical regime
 simply by varying the sweep speed.

\begin{figure}[h] 
\epsfig{file=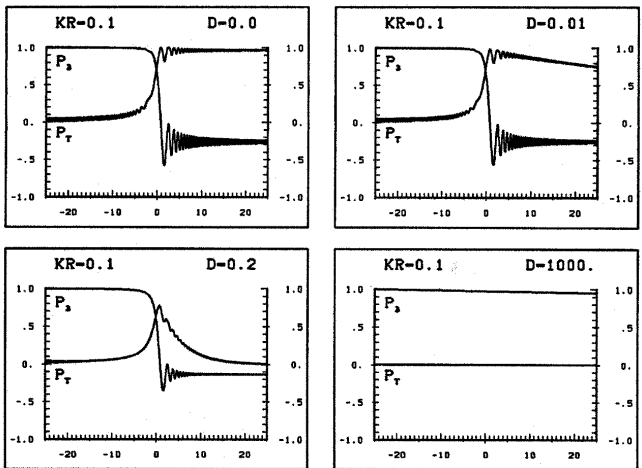, width=\hsize}
\caption{ Same as Fig. 5, but with a faster sweep (smaller $KR$),
showing the effects of a mild non-adiabaticity. $\bf P$ fails to
follow $\bf V$. When there is no damping $\bf P$ remains with wide 
oscillations. With moderate damping $\bf P$ gradually shrinks to
zero. Finally with very strong damping the  ``Turing-Watched Pot-
Zeno''behavior  keeps $\bf P$ essentially constant.} \end{figure}

\section{Parameter Values}
 
We now discuss some some typical parameters of our system. In
particular since we
 envision working at time scales shorter than have been common in
this field, 
it may be useful to give a qualitative discussion of the various
time scales involved.
 The highest
frequency or shortest time present is the ordinary harmonic
frequency
$\omega_0$ (Fig. 1)   giving
 the approximate level spacing 
  in the well. For typical conditions this spacing may be on the
order
of
 several Kelvin  ($K$), or some hundred  $GHZ$. Since we
suppose working
 in the Kelvin to milliKelvin range, this implies that for the
equilibrium
system at the start of the sweep
 only the lowest level is populated.
 The next parameter is the tunneling
frequency through the barrier.
 This is a sensitive function of the SQUID
parameters, with sample values [$\beta_L= 1.19,~ C=0.1~pF,
 L= 400~pH, ~ R=5~M\Omega$] we have $\omega_{tunnel} \approx
600~MHZ$. This corresponds to $\omega_{tunnel}/\omega_0\sim 10^{-
2}$,  the tunneling introduces small energy shifts on the scale of
the principal level splitting

 The
 next two parameters concern the extent and speed of the sweep.  
The initial asymmetry  $\epsilon$ (Fig. 1) should
be small compared to $\omega_{0}$ in order to retain
 the approximate two-state character of the system, but large
compared to
 $\omega_{tunnel}$ to avoid initial mixing of  the two states,  say
$\epsilon/ \omega_{0}\sim 10^{-1}-10^{-2}$. The speed of the sweep,
$\omega_{sweep}$ will be the easiest  experimental
 parameter to control, and the behavior of the results as
$\omega_{sweep}$ is varied will be an 
important check on the theory. It must not be so fast
as to
lose adiabaticity, but not so slow as to allow
relaxation processes to mask
the results. We may suppose it to be in the range
 $10^{-3}\omega_{0}-10^{-4}\omega_{0}= 10-100~MHZ $. 
With these numbers it seems  possible to satisfy the
adiabatic condition Eq~[\ref{adb}]on the one hand and to sweep fast
relative to the relaxation time (see below) on the other.

 We now come to the
dissipative parameters: $D$,
 and $\omega_{relax}$, the
relaxation
rate for transitions among the SQUID levels. The calculation of
dissipative effects in the SQUID is usually approached in terms of
the 
Caldeira- Leggett model ~\cite{cl}, where a coupling to a
pseudoboson field, related to the resistance of the device, 
represents the dissipative effects. The distinction between
relaxation and decoherence in the SQUID is principally a question
of energy scale. For the  former, influences (e.g. the
pseudobosons), involving jumps between levels, energies on the
order of the level splitting or more are involved. For the latter,
where there is only a ``dephasing'',  low energies, those below the
level splitting,  are important. In general, of course, both
processes are present, but at low enough temperture relaxation
should become small. For this reason the damping term
in~Eq~[\ref{pdot}] is taken to only affect the transverse
components of $\bf P$.  

 Weiss and Grabert~\cite{wg}, have given a calculation of the
effects of dissipation on coherence, and we may identify their
``decay rate'' at
weak dissipation with  $D$. This gives the estimate    $D=T/Re^2$
(their $\Gamma$). 
For
$R=5~M\Omega$,  we find $D= 0.08~mk=9.6~MHZ$  at $T=100~mk$, and
$D=
0.008~mk=960~kHZ$  for $T=10~mK$. (Units: $1~K= 8~meV=120~GHZ$,
$1/e^2=4~k\Omega$). With these estimates, say 1-10~MHZ,
we are
in an interesting range  since
  as mentioned, this offers the
interesting
possiblity of being able to choose  sweep speeds either slow or
fast
with respect to the decoherence time,  while still retaining the
adiabatic condition.
The resulting switches between
classical and  quantum behavior would provide persuasive evidence
for the correctness of our general picture,
 and allow the measurement of $D$ and  its temperature dependence. 

 Our last parameter is the
relaxation
 rate  $\omega_{relax}$, characterizing, as said, the rate of
conventional kinetic
 relaxation processes. We wish this to be small for two reasons.
One is
that relaxation should not take place during the sweep, which would
obviously obscure our effects. Secondly, relaxation should also be
small during the readout time, otherwise for example, the final
configuration of Fig. 4 might turn into that of Fig. 3 before it
could be
detected.

Here we may use calculations which have been made in connection
with the tunneling to the voltage state ~\cite{ovi}. The rate one
finds depends on the separation of the states in question. If we
call 
 $\omega_{relax}$ the value with the level seperation at the
beginning or end of the sweep we obtain in our example$~ \sim
20~kHZ$. This characterizes the time in which the readout must
occur.

 During the sweep itself the level spacing is changing. Taking this
into account we obtain the curve of Fig. 7, which shows the
probability of an inter-level jump as a function of
$~\omega_{sweep}$. It appears that for sweeps shorter than a
microsecond relaxation is indeed negligible.  

To indicate the  general relationship of the various time scales,
the   frequencies  $\omega_{relax}, \omega_{dec}=1/D$ and $~
\omega_{tunnel}$ are marked for the horizontal axis. As may be
seen,  we may indeed be able to obtain the favorable situation  of
being able to sweep slow or fast relative to the decoherence time,
while retaining small relaxation. 

\begin{figure}[h] 
\epsfig{file=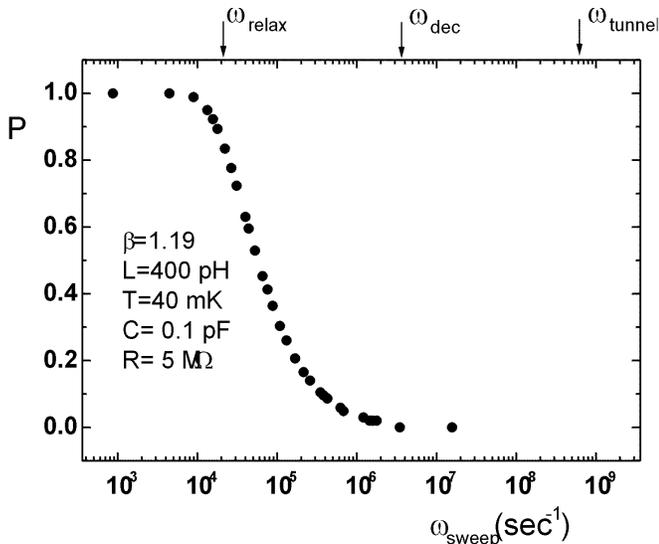,width=\hsize}
\caption{ The curve shows the probability of relaxation during a
sweep, as a function of the sweep frequency (inverse sweep time).
Important time scales or frequencies are marked for the parameters
indicated. It appears possible to have sweeps either slow or fast
relative with respect to the decoherence time, while retaining
small relaxation.} \end{figure}

Concerning the readout,
 we have examined a scheme involving
  a switchable flux linkage to a DC SQUID,
 which would  be sufficiently fast for the above
estimates~\cite{dino}.
 Here one profits from the fact that observation
 of the system is only necessary after the procedure is completed.

\section{Nonadiabatic Procedures}
 In general, we need not necessarily limit ourselves to slow,
adiabatic processes. There is, say,  the opposite case of the
``sudden approximation'' where $\bf V$ is changed very quickly and
$\bf P$ tends to stays put, as was beginning to happen in Fig. 6.
This could
be of interest  for example, if we find experimentally that we
always get  an inversion (Fig. 3), and never a failed inversion
(Fig. 4)  for the range of parameters available to us
operationally. There could be two reasons for this behavior. 
It could be a true result in the sense that the decoherence is very
small; we are always in the quantum situation and the adiabatic
inversion always works.
 
But we might worry that instead we have a rapid relaxation from the
upper state to the lower one after or during the sweep. That is,
perhaps we really had no inversion and Fig. 4 applies. The
relaxation rate is greater than we think and the system just falls
back to the lowest energy state with the emission of some energy
before we can read out. 

A way to clarify the situation would  be to  start our adiabatic
inversion procedure from the {\it upper} state.  Then a succesful
inversion means we  end in the upper state. If relaxation  was not
the problem,  we should then detect the upper state. On the other
hand if relaxation is significant, we should always end in the  
lower state, regardless of where we start.

Since at low temperature everything is in the lowest state, in
order perform  this check we need a method of getting the system
into the
upper state at the start of the sweep. We can accomplish this by a 
sudden, nonadiabatic, reversal of $V_{vertical}$, that is, of  the
applied flux. Since the wavefunction or $\bf P$ will 
(approximately)  not change, we will have the system in the upper
state to start with. We now proceed with the adiabatic sweep as
before and we can check if the results were due to relaxation or
true quantum coherence.
 
Other interesting configurations and procedures can undoubtedly be
arrived at by combining various operations in this way.

\section{qbits and the quantum computer}  The two-state system
under discussion here suggests
itself as a physical embodiment of  the ``quantum computer''.   
The ``qbit'' itself is naturally represented by
L and R playing the role of 0 and 1.  A  linear
combination of L and R may be created by adiabatically rotating
$\bf P$
from some starting position. 
 Adiabatic inversion is evidently an embodiment of NOT since it 
will turn one linear combination into another one with the weights
of L and R interchanged.
 As for  CNOT, the other basic operation,  a NOT is performed  or
not performed  on a
``target bit'' according to the state of a second, ``control bit'',
which itself does not change its state.
  One straightforward realization of this would be to
 perform the NOT operation as just described in the presence of an
additional linking flux
supplied by a  second  SQUID nearby.  The magnitude and direction
of this linking flux  would
be so arranged that the inversion of the first SQUID is or is not
successful depending on the state
of the second SQUID. This and many other interesting combinations
of junctions and flux
linkages, not to mention other devices~\cite{averin}, may be
contemplated and are under study~\cite{cnot}.

SQUID systems like these would seem to be particularly well suited 
for the  embodiment of the quantum computer, where we  wish to
generate a series of unitary
transformations for  the various steps of computation.
 This may be done by creating 
a ``moving landscape" of potential maxima and minima, as in our
simplest
one-dimensional example of the adiabatic inversion. This imaginary
landscape 
  can be produced and manipulated by controlling various external
parameters
 (as with our sweeping flux), performing the various 
operations in one physical device. Naturally, practicality
will depend very much on the relation between the speed of these
operations and the decoherence/ relaxation times which we hope to 
determine by the present methods.

Although our interest here has been the SQUID, it will be evident
that the principle of determining decoherence times through the
inhibition of adiabatic inversion could be applied to many other
types of systems as well.


\end{document}